\documentclass{XrU2005}

\usepackage{psfig,color,pstricks}
\title{XMM-{\it Newton} RGS and {\it Chandra} LETGS observations of
the WHIM in 1ES 1028+511}
\author{K. C. Steenbrugge}
\author{F. Nicastro}
\author{M. Elvis}
\affil{Harvard-Smithsonian Center for Astrophysics, 60 Garden street,
Cambridge, MA 02138, USA}

\begin{document}


\maketitle

\begin{abstract}
We report preliminary results on the detection of the Warm-Hot
Intergalactic Medium (WHIM) along the line of sight toward the blazar
1ES 1028+511 (z = 0.361). 1ES 1028+511 was observed for 150~ks with the low  energy transmission 
grating in combination with the high resolution camera onboard
{\it Chandra}. An additional 300 ks observation was obtained using the
reflection grating spectrometers (RGS) onboard XMM-{\it Newton}. 
We report the detection of three absorption lines which can be
attributed to the 
WHIM, and compare the results with theoretical predictions. 
\end{abstract}

\section{Introduction}
At z $>$ 2 the vast majority of the baryons ($>$ 76 $\%$,
\citet{rauch98}; \citet{weinberg97}) are found in a 
mildly photoionized (by the metagalactic UV radiation field) phase of 
the IGM, through a forest of HI Ly$\alpha$ absorption lines in the 
background O-UV spectra of background quasars. However, at z $<$ 2 
only $\sim$ 30 $\%$ of the baryons are detected in the residual local 
Ly$\alpha$ forest \citep{penton04}, and 
already-virialized structures account for even a smaller fraction $\sim 
12$ \% \citet{fukugita03}. About 54 \% of the 
baryons are eluding detection in the local Universe. Hydrodynamical 
simulations for the formation of structures in a $\Lambda$-CDM Universe, 
provide a self-consistent solution to this puzzle: about half of the 
baryons in the local Universe should still be confined in the IGM, but 
shock-heated to temperatures of about $10^5-10^7$ K, during the collapse 
of density perturbation.

At such high temperatures H is fully ionized and so the gas is 
transparent to Optical and UV observations. However, electronic 
transitions from highly ionized metals, can still provide a significant 
source of opacity. CV-VI, OVII-VIII and NeIX-X K$\alpha$ are the most 
intense of these transitions, in gas with Solar-like composition. All 
these transitions fall in the soft X-ray band, and can now be detected 
thanks to the high-resolution spectrometers of {\it Chandra} and 
XMM-{\it Newton}.

1ES 1028+511 (\cite{schachter93}; \cite{elvis92}) is at a redshift
of 0.361 \citep{polomski97} thus, according to simulations
\citep{fang02}, we expect to detect 1 system with an O~VII column
density equal or greater than 10$^{16}$ cm$^{-2}$.

\section{Observations and data reduction}
1ES~1028+511 was observed by {\it Chandra} for 149.86 ks on March 11, 2004, using
the Low Energy Transmission Grating in combination with the
High Resolution Camera (LETGS). 
We extracted the LETGS spectrum using the pipeline described by \citet{kaastra02}, which includes an empirical correction for the 
known wavelength problem in the LETGS \citep{kaastra02} and fitted it with responses that include the 
first 10 positive and negative orders.

The XMM-{\it Newton} observation was split into three separate
observations, between June 20 to 24, 2005. The exposure times were
104.2, 95.2 and 101.4 ks. 
The data were reduced using the SAS version
6.1.0 standard threads. 

In all fitting, we fitted the RGS and LETGS spectra (7 spectra) 
simultaneously. Errors were evaluated at 68 $\%$ significance level, for 
one interesting parameter. The data were analyzed using the {\it spex} package
\citep{kaastra02b}. 

\section{Absorption lines}
Three absorption lines are detected with a at least 2 $\sigma$ significance. 
Table 1 lists the equivalent width (EW) of these absorption features,
the most likely identification, and the ionic column
densities assuming a velocity broadening of 100 km s$^{-1}$. 

\begin{table}
\begin{center}
\caption{The observed wavelength, EW, ion column density, the redshift
and the most likely identifications.}
\begin{tabular}{l|l|l|l|l} \\
$\lambda$$_{obs}$       & EW      & log N$_{\rm i}$ & redshift & iden. \\
 \AA    & m\AA          & cm$^{-2}$ &      &       \\\hline
28.74   & 34$\pm$12     & 15.7$\pm$0.4 & 0.33    & O VII K$\alpha$   \\
46.25   & 24$\pm$7    & 15.6$\pm$0.2   & 0.15     & C V    K$\alpha$    \\
        &               &         & 0.12   & C IV   K$\alpha$    \\
48.80   & 32$\pm$10   & 15.8$\pm$0.2   & 0.21   & C V    K$\alpha$    \\
        &               &         & 0.18   & C IV   K$\alpha$    \\
\end{tabular} \\
\end{center}
\end{table}
The 28.74~\AA~line can be identified as an O VII K$\alpha$ line and
has a 2.5 $\sigma$ significance. However, this line is only seen in the positive
order of the LETGS and has a full width half maximum of 500 km s$^{-1}$. The line falls on an instrumental feature
in the RGS spectra. An alternative, but unlikely identification is 
z=0 N~VI K$\alpha$. Therefore it should be considered as a possible
detection of O VII K$\alpha$. 

The 46.25~\AA~and 48.80~\AA~lines can be either
identified as C~V or C~IV lines. The significance of the lines is 3.4
and 3.2 $\sigma$ respectively. Table 2 lists the OVII$-$OVIII column density 
1-sigma upper limit at the redshifts of the CV identifications. If the lines are C~IV, they are 
probably imprinted by mildly photoionized gas at temperatures of about 
$10^4$ K, and so are not tracking WHIM filaments. For both
redshifts the C~IV 1549~\AA~line is redshifted out the HST STIS
spectrum. 

\begin{table}
\begin{center}
\caption{The 1 $\sigma$ upper limits for the O~VII and O~VIII column density.}
\begin{tabular}{l|l|l} \\
redshift    & N$_{O VII}$  & N$_{O VIII}$ \\
            & log cm$^{-2}$& log cm$^{-2}$ \\\hline
0.15       & $>$ 15.5     & $>$ 15.9          \\
0.21       & $>$ 15.3     & $>$ 15.9          \\
\end{tabular}
\end{center} 
\end{table}

\section{Comparison with simulation}

\begin{figure}
\centering
\psfig{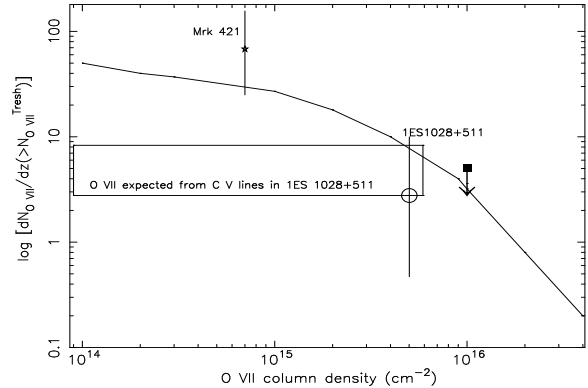}
\caption{The predicted number of intervening
WHIM absorbers per redshift versus the O VII column density (solid line)
\citep{fang02}. The result for the Mrk~421 spectra \citep{nicastro05}
(star), the Gehrels upper
limit (square), and assuming we detect O~VII (circle). The box indicates the expected number of O~VII
absorption lines and their column density calculated from the C~V
detections in the 1ES~1028+511 spectra.}
\label{fig:sim}
\end{figure}

Fig.~\ref{fig:sim} shows the expected number of WHIM absorbing systems
per unit redshift versus O VII column density \citep{fang02}. The the
Gehrels upper limit (square), the measured column density for O VII
(circle), and the measured column
density for Mrk 421 (star). The box indicates the expected number of O~VII
lines and column densities from the detected C~V lines. All
data points are consistent with the expectation of hydrodynamical
simulations by \citet{fang02}. 

\section{Acknowledgements}
K. C. S. and F. N. acknowledges support from the Chandra grant
G04-5101X. F. N. acknowledges support from the XMM grant NNN046D836.

\bibliography{references}

\begin{thebibliography}{11}
\expandafter\ifx\csname natexlab\endcsname\relax\def\natexlab#1{#1}\fi

\bibitem[{{Elvis} {et~al.}(1992){Elvis}, {Plummer}, {Schachter}, \&
  {Fabbiano}}]{elvis92}
{Elvis}, M., {Plummer}, D., {Schachter}, J., \& {Fabbiano}, G. 1992, \apjs, 80,
  257

\bibitem[{{Fang} {et~al.}(2002){Fang}, {Bryan}, \& {Canizares}}]{fang02}
{Fang}, T., {Bryan}, G.~L., \& {Canizares}, C.~R. 2002, \apj, 564, 604

\bibitem[{Fukugita(2003)}]{fukugita03}
Fukugita, M. 2003, Cosmic Matter Distribution: Cosmic Baryon Budget Revisited

\bibitem[{{Kaastra} {et~al.}(2002b){Kaastra}, {Mewe}, \&
  {Raassen}}]{kaastra02b}
{Kaastra}, J.~S., {Mewe}, R., \& {Raassen}, A.~J.~J. 2002b, Proc. Symp. New
  Visions of the X-ray Universe in the XMM-{\it Newton} and {\it Chandra} era

\bibitem[{{Kaastra} {et~al.}(2002){Kaastra}, {Steenbrugge}, {Raassen}, {van der
  Meer}, {Brinkman}, {Liedahl}, {Behar}, \& {de Rosa}}]{kaastra02}
{Kaastra}, J.~S., {Steenbrugge}, K.~C., {Raassen}, A.~J.~J., {et~al.} 2002,
  \aap, 386, 427

\bibitem[{{Nicastro} {et~al.}(2005){Nicastro}, {Mathur}, {Elvis}, {Drake},
  {Fiore}, {Fang}, {Fruscione}, {Krongold}, {Marshall}, \&
  {Williams}}]{nicastro05}
{Nicastro}, F., {Mathur}, S., {Elvis}, M., {et~al.} 2005, \apj, 629, 700

\bibitem[{{Penton} {et~al.}(2004){Penton}, {Stocke}, \& {Shull}}]{penton04}
{Penton}, S.~V., {Stocke}, J.~T., \& {Shull}, J.~M. 2004, American Astronomical
  Society Meeting Abstracts, 204,

\bibitem[{{Polomski} {et~al.}(1997){Polomski}, {Vennes}, {Thorstensen},
  {Mathioudakis}, \& {Falco}}]{polomski97}
{Polomski}, E., {Vennes}, S., {Thorstensen}, J.~R., {Mathioudakis}, M., \&
  {Falco}, E.~E. 1997, \apj, 486, 179

\bibitem[{{Rauch}(1998)}]{rauch98}
{Rauch}, M. 1998, \araa, 36, 267

\bibitem[{{Schachter} {et~al.}(1993){Schachter}, {Stocke}, {Perlman}, {Elvis},
  {Remillard}, {Granados}, {Luu}, {Huchra}, {Humphreys}, {Urry}, \&
  {Wallin}}]{schachter93}
{Schachter}, J.~F., {Stocke}, J.~T., {Perlman}, E., {et~al.} 1993, \apj, 412,
  541

\bibitem[{{Weinberg} {et~al.}(1997){Weinberg}, {Miralda-Escude}, {Hernquist},
  \& {Katz}}]{weinberg97}
{Weinberg}, D.~H., {Miralda-Escude}, J., {Hernquist}, L., \& {Katz}, N. 1997,
  \apj, 490, 564

\end{thebibliography}

\end{document}